\documentclass[%
 reprint,
nofootinbib,
nobibnotes,
 amsmath,amssymb,
 aps,
]{revtex4-1}

\usepackage{aas_macros}
\usepackage{float}
\usepackage{graphicx}
\usepackage{dcolumn}
\usepackage{bm}
\usepackage{hyperref}

\newcommand{\EE}{$\Omega_{\rm ee}$~}

\begin{document}

\preprint{APS/123-QED}

\title{Early dark energy, the Hubble-parameter tension, and the string axiverse}

\author{Tanvi Karwal}
\author{Marc Kamionkowski}%
\affiliation{%
 \mbox{Department of Physics and Astronomy, Johns Hopkins University,} 
 \\3400 N. Charles St., Baltimore, MD 21218
}%

\date{\today}

\begin{abstract}

	Precise measurements of the cosmic microwave background (CMB) power spectrum are in excellent agreement with the predictions of the standard $\Lambda$CDM cosmological model.  However, there is some tension between the value of the Hubble parameter $H_0$ inferred from the CMB and that inferred from observations of the Universe at lower redshifts, and the unusually small value of the  dark-energy density is a puzzling ingredient of the model.  In this paper, we explore a scenario with a new exotic energy density that behaves like a cosmological constant at early times and then decays quickly at some critical redshift $z_c$.  An exotic energy density like this is motivated by some string-axiverse-inspired scenarios for dark energy.  By increasing the expansion rate at early times, the very precisely determined angular scale of the sound horizon at  decoupling can be preserved with a larger Hubble constant.  We find, however, that the Planck temperature power spectrum tightly constrains the magnitude of the early dark-energy density and thus any shift in the Hubble constant obtained from the CMB.  If the reionization optical depth is required to be smaller than the Planck 2016 $2\sigma$ upper bound $\tau\lesssim 0.0774$, then early dark energy allows a Hubble-parameter shift of at most 1.6 km~s$^{-1}$~Mpc$^{-1}$ (at $z_c\simeq 1585$), too small to fully alleviate the Hubble-parameter tension.  Only if $\tau$ is increased by more than $5\sigma$ can the CMB Hubble parameter be brought into agreement with that from local measurements.  In the process, we derive strong constraints to the contribution of early dark energy at the time of recombination---it can never exceed $\sim2\%$ of the radiation/matter density for $10 \lesssim z_c \lesssim 10^5$. 



\end{abstract}

\maketitle


\section{Introduction} \label{sec_intro}

	Current measurements of temperature and polarization power spectra of the cosmic microwave background (CMB) are in excellent agreement with the standard $\Lambda$CDM cosmological model \cite{Ade:2013zuv}.  Still, there is some tension between the value $H_0 = 66.93 \pm 0.62$ km~s$^{-1}$~Mpc$^{-1}$ of the Hubble parameter obtained from the CMB \cite{Aghanim:2016yuo} and those obtained from local measurements, $H_0 = 73.24 \pm 1.74$  km~s$^{-1}$~Mpc$^{-1}$ ($3.4\sigma$ tension) as inferred from supernovae and more \cite{Riess:2016jrr}, and $H_0 = 72.8 \pm 2.4$  km~s$^{-1}$~Mpc$^{-1}$ as measured by H0LiCOW \cite{Bonvin:2016crt}.  There is also unease among some theorists about the incredibly small value, relative to the Planck density, of the dark-energy density required to account for the observations \cite{Caldwell:2009ix}.  There are an almost endless number of explanations for dark energy, but this work will be inspired by a recently proposed string-axiverse \cite{Svrcek:2006yi,Svrcek:2006hf,Arvanitaki:2009fg,Marsh:2011gr} scenario for dark energy \cite{Kamionkowski:2014zda}.

	The purpose of this paper is to investigate whether the Hubble-parameter tension might be explained by the presence of an exotic dark-energy density in the early Universe of the type that might arise in some of these axiverse scenarios.  In this framework, dark energy is due to an axion-like field that is active today \cite{Marsh:2011gr,Kamionkowski:2014zda}.  However, there can be a large number of similar light fields that can be dynamically important at some point in the earlier history of the Universe and then decay away in influence.

	Here we surmise that one of these axion-like fields becomes dynamical around the time of recombination.  More precisely, it behaves, as we will delineate more clearly below, like a cosmological constant at early times.  However, at some critical redshift $z_c$, which is taken to be on order the redshift of recombination, the energy density then decays more rapidly than that of radiation.  The cosmological-constant--like behavior at early times increases the pre-recombination expansion rate and thus reduces the sound horizon at recombination.  The resulting reduction in the angle subtended by the CMB acoustic peaks can then be compensated by an increase in the Hubble constant. 

	Although such an exotic early dark energy is capable of increasing the value of the Hubble parameter today, we find that a value of $\tau$ greater than its Planck-2016 $2\sigma$ upper bound is required to fully resolve the Hubble tension. We also find that the exotic energy (EE) is constrained to contribute at most $\sim 2\%$ of the total energy density of a $\Lambda$CDM universe around the time of recombination, and may only contribute $\gtrsim 5\%$ if it decays earlier than a redshift of $\sim 10^5$. 

	The idea of an additional early-Universe contribution to the energy density has been considered before \cite{Doran:2006kp, Pettorino:2013ia, Calabrese:2011hg}. Although similar in spirit, those models differ from what we consider here.  The conclusion reached in our work that EE contributes no more than $\sim 2\%$ of the critical density around the time of recombination is consistent with the conclusions of earlier papers on other early-dark-energy models in which upper limits of $\sim 4-5\%$ were inferred. The increased early-Universe expansion rate considered here also resembles in spirit the explanation suggested in Refs.~\cite{Riess:2016jrr, Calabrese:2011hg, Bernal:2016gxb}  for the Hubble-parameter tension in terms of an increased number of relativistic degrees of freedom.

	This paper is organized as follows. In Section \ref{sec_model}, we describe the exotic energy model, its evolution and its effect on the TT spectrum. In Section \ref{sec_method}, we describe the Fisher-matrix analysis we employ to constrain the model (Section \ref{sec_Fish_partials}) and the data we use for this analysis (Section \ref{Sec_Pl_data}). In Section \ref{sec_constraints} we obtain constraints on the EE and determine how it changes the Hubble parameter. We do so for the optical depth at reionisation $\tau$ fixed at its current best-fit (Section \ref{sec_Pl_tau}), $2\sigma$ (Section \ref{sec_2sig}) and $5\sigma$ (Section \ref{sec_5sig}) values. We conclude in Section \ref{sec_conclusion}.

\section{Model} \label{sec_model}

	The form of the exotic energy (EE) we consider is motivated by the axion-like fields discussed in Ref.~\cite{Kamionkowski:2014zda}.  There it was argued that an axion-like field driving accelerated expansion today might be one of $\sim100$ such fields in the string axiverse, each of which has some small chance to drive accelerated expansion at some point in the history of the Universe.  The scenario suggests that there may be other axion-like fields that may have behaved earlier in the history of the Universe like a cosmological constant but then decayed away in influence.

	Here we will use a phenomenological model inspired by Ref.~\cite{Kamionkowski:2014zda}.  The energy density $\rho_{\rm ee}$ of the EE takes the form,

	\begin{equation}
		\frac{\rho_{\rm ee}(a)}{\rho_c} = \frac{\Omega_{\rm ee} (1 + a_c^6)}{a^6 + a_c^6} ,
	\end{equation} \label{rho_EE}
	where $\rho_c$ is the critical density today, \EE is the fractional energy density of the EE today and $a_c = 1/(1+z_c)$ is the critical value of the scale factor at which the EE shifts from early-time behavior to late-time behavior. 

	The pressure the EE exerts is 
	\begin{equation}
		p_{\rm ee}(a) = \rho_{\rm ee} \frac{a^6 - a_c^6}{a^6 + a_c^6} .
	\end{equation} \label{pressure_EE}
	It can be seen that at redshifts $z \gg z_c$, we have $a^6 \ll a_c^6$ and therefore $p_{ee} \simeq - \rho_{ee}$. That is, the EE behaves like a cosmological constant at early times, similar to a slowly rolling axion field. On the other hand, at redshifts $z \ll z_c$, $a^6 \gg a_c^6$ and $p_{ee} \simeq \rho_{ee}$, emulating a free scalar field, with the hardest possible equation of state allowed by causality. 

	\begin{figure}[h]
		\includegraphics[width = 0.5\textwidth]{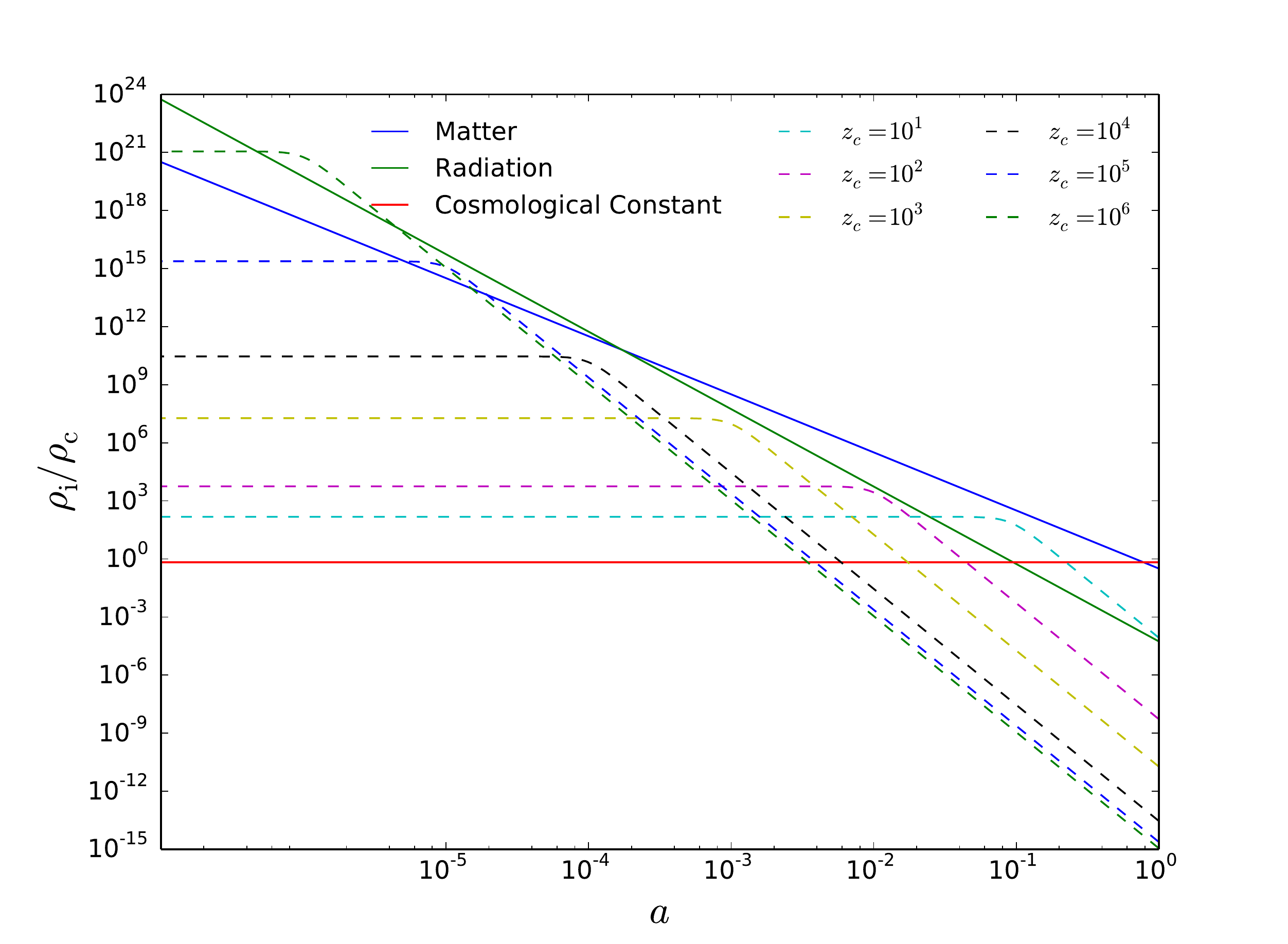}
		\caption{Shown here are the evolutions of the energy densities of exotic energy (EE; dashed lines) for several critical redshifts $z_c$, matter (solid blue), radiation (solid green), and the cosmological constant (solid red).  For each $z_c$ we choose the exotic-energy density \EE to be the $3\sigma$ upper limit we derive from the Planck temperature power spectrum assuming the reionization optical depth $\tau$ is fixed to the current Planck best-fit value.  The energy densities are all shown, relative to the critical density $\rho_c$ today, as a function of the scale factor $a$. }
		\label{EE_behaviour}
	\end{figure}

	Fig.~\ref{EE_behaviour} shows how the energy density of the EE evolves over cosmic history. Matter, radiation and the cosmological constant are also shown for comparison. Changing \EE shifts the curve of the EE up or down. Changing $z_c$ changes the redshift at which the EE switches from behaving like a cosmological constant to decaying away faster than radiation. 

	We assume that the EE only changes the homogeneous background evolution of the universe. We do not have a physical model of how perturbations change as a result of adding this phenomenological model to $\Lambda$CDM. In this paper, we simply add the energy density and pressure of the EE to the Friedmann equation in the background sector of the public code Cosmic Linear Anisotropy Solving System (CLASS) \cite{Blas:2011rf}. We note that inclusion of scalar-field perturbations can, in some cases, considerably alter the perturbation spectrum \cite{Caldwell:1997ii}. We will address the effects of perturbations in realistic, physical models for EE in subsequent work. 

	\begin{figure}[h]
		\includegraphics[width = 0.475\textwidth]{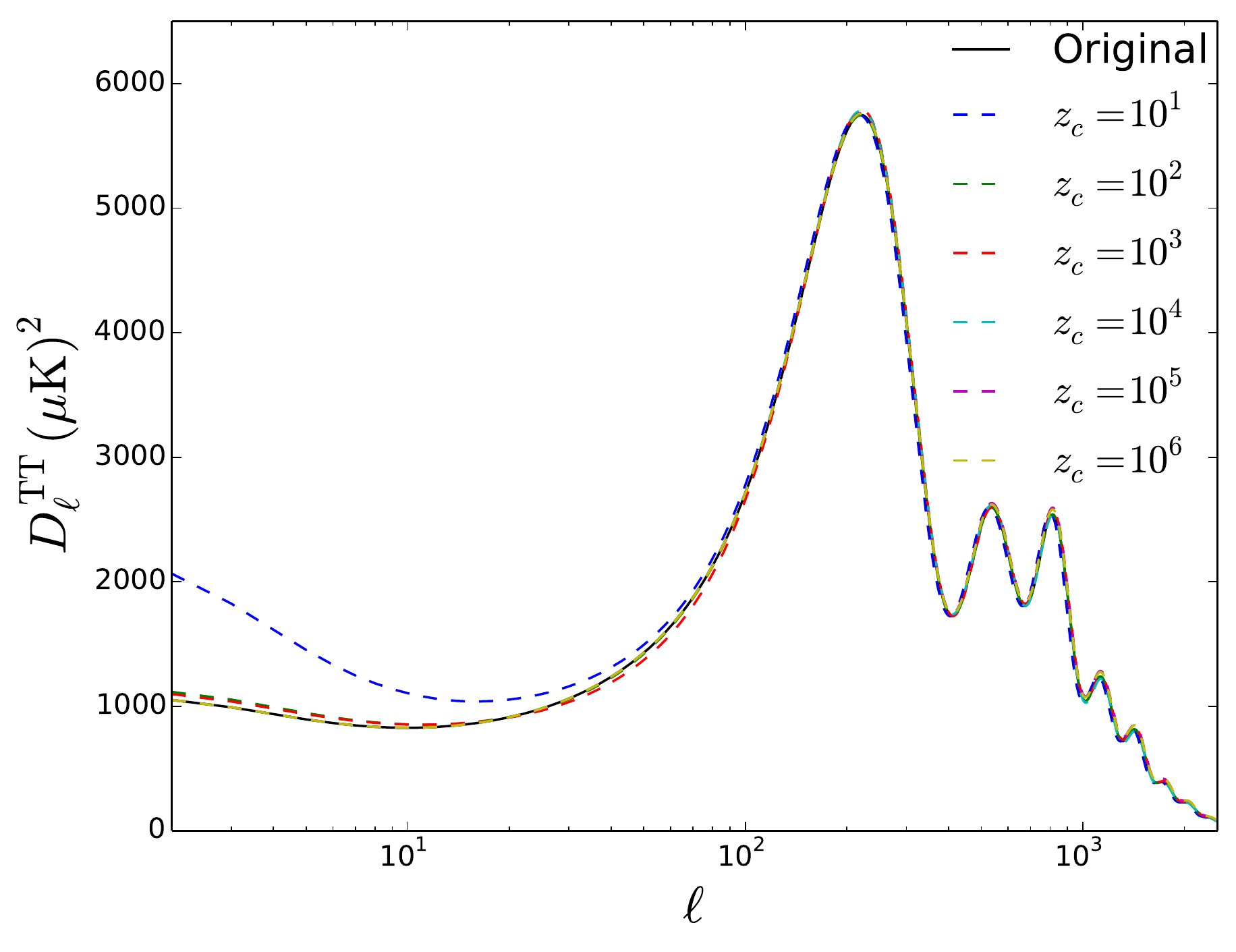}
		\includegraphics[width = 0.475\textwidth]{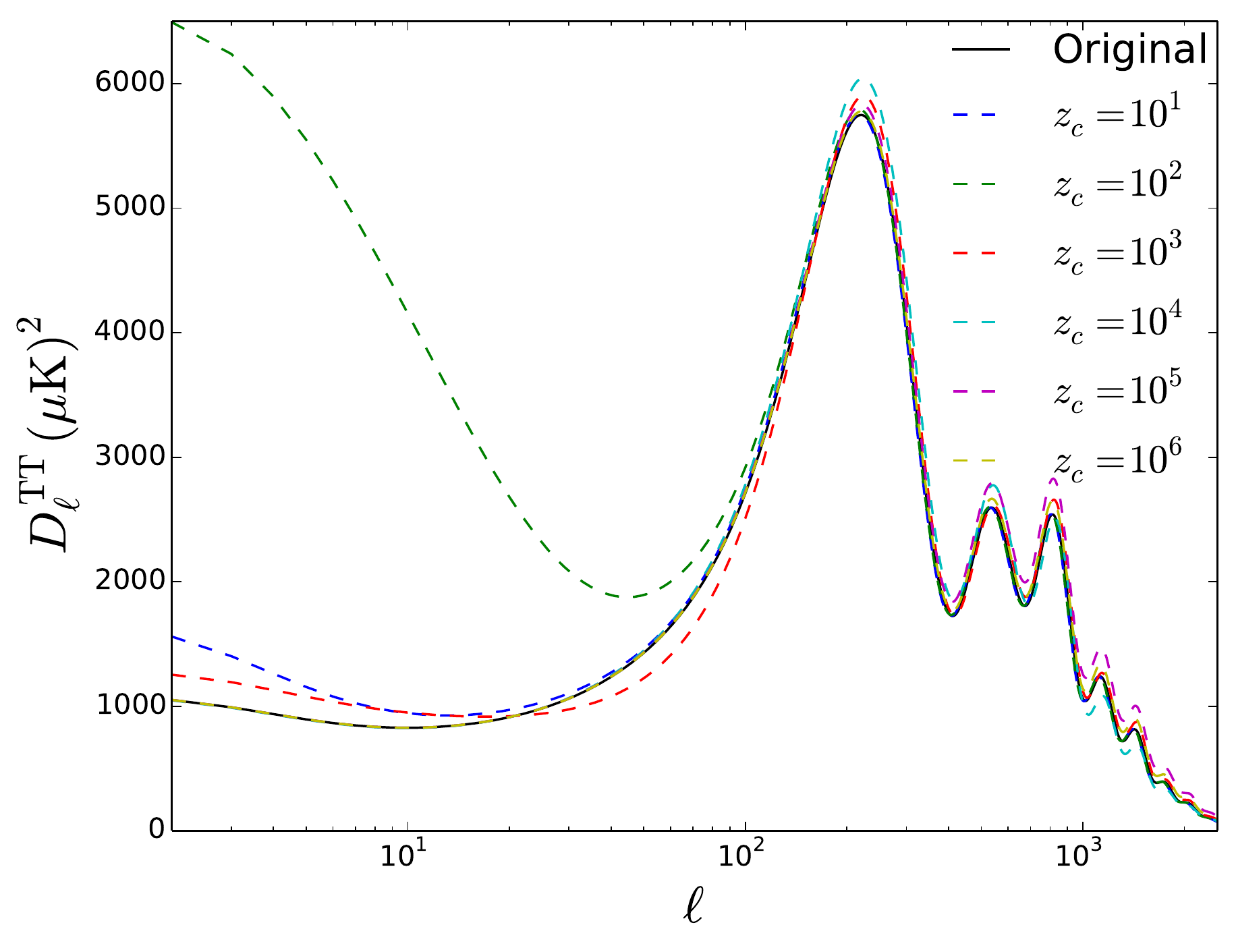}
		\caption{The shifts caused in the TT power spectrum due to the addition of EE are shown for various $z_c$.  Here, the value of the reionization optical is fixed to the current Planck best-fit value $\tau=0.0596$.  The other cosmological parameters are fixed at the values, shown in Table I, that provide the best fit to the TT power spectrum.  Clearly the critical redshift of the EE is important in determining how the EE shifts the TT spectrum. 
		In the upper figure, the value of \EE chosen for each $z_c$ is the $3\sigma$ upper limit of its best fit. 
		In the lower figure, \EE is chosen such that it moves $\theta_*$ by 1\%. \EE is approximately two orders of magnitude greater for the lower plot. }
		\label{EE_Shifted_spectra}
	\end{figure}

	On adding such an EE with non-zero \EE to $\Lambda$CDM, the predicted TT angular power spectrum will shift. It can be shifted back to better fit the data by shifting the other parameters of the $\Lambda$CDM model. We show how EEs of various $z_c$ and \EE shift the TT spectrum in Fig.~\ref{EE_Shifted_spectra}. 

	For our analysis, we choose the critical redshift range $10 \leq z_c \leq 10^6$. We found that critical redshifts smaller than approximately 500 shift the angular size $\theta_*$ of the sound horizon at recombination to larger values for \EE$>0$. If $\theta_*$ were increased in this way, then the current expansion rate $H_0$ would have to be decreased to shift $\theta_*$ back to its measured value. EEs with $z_c\lesssim 500$ therefore move the Hubble parameter further away from its local value, exacerbating the discrepancy between the Planck and local values. We include some such critical redshifts in our analysis, limiting the $z_c$ range to 10 on the lower end. 

	On the higher end, we limit our analysis to $z_c \leq 10^6$, as EEs with higher critical redshifts have little effect on CMB power spectra.

\section{Method} \label{sec_method}

	Our aim here is to determine the largest value of the fractional exotic energy density \EE consistent with Planck measurements of the temperature power spectrum\footnote{Observations obtained with Planck (http://www.esa.int/Planck), an ESA science mission with instruments and contributions directly funded by ESA Member States, NASA, and Canada.}, after marginalizing over the other cosmological parameters that are fit to the data. (While doing so, we also investigate whether a nonzero \EE is preferred by the data, but find a null result.)  Given the speculative nature of the model, here we do a rough initial analysis, following that outlined in Refs.~\cite{Jungman:1995bz,Tegmark:1996bz,Galli:2014kla,Munoz:2015fdv}, in which the log-likelihood is approximated by a quadratic dependence on the parameters.  The loss of precision of this approach, relative to the full Monte Carlo analysis, is made up for by clarity and simplicity.  The upper bounds we derive, though, should be understood as approximations rather than precise results.

	Given the complexities involved in the current Planck polarization data, we work here with only the temperature power spectrum.  Since the primary impact of the polarization data (especially that at low multipole moments $\ell$) is to fix the reionization optical depth $\tau$ \cite{Zaldarriaga:1996ke}, we remove $\tau$ from our Fisher analysis and instead fix it to different values that fall within (and, for illustration, also outside) the current Planck error limits.  As we will see, the best-fit cosmological parameters we infer from the temperature power spectrum are in rough agreement (within $2\sigma$) of those reported by the complete Planck analysis (including polarization).  We believe, therefore, that the cosmological-parameter shifts we infer below from the introduction of exotic energy reflect reasonably well those that would be obtained from a complete analysis.

	\subsection{Fisher Matrices} \label{sec_Fish_partials}

		In order to constrain \EE for various $z_c$'s, we do a Fisher-matrix analysis using the Planck TT angular power spectrum $D_{\ell}^{\rm TT,obs}$ in a manner similar to that outlined in Ref.~\cite{Jungman:1995bz,Tegmark:1996bz,Galli:2014kla,Munoz:2015fdv}. For the analysis, we vary $H_0 = 100h$ km s$^{-1}$ Mpc$^{-1}$, the fractional density $\omega_{\rm b} = \Omega_{\rm b} h^2$ of baryons today, the fractional density $\omega_{\rm c} = \Omega_{\rm c} h^2$ of cold dark matter today, the amplitude ln$(10^{10}A_{\rm s})$ of the primordial power spectrum, and the scalar spectral index $n_{\rm s}$. We refer henceforth to these 5 parameters in our Fisher analysis as the ``cosmological parameter'' and then introduce the current exotic-energy density \EE, for a given $z_c$, as a sixth parameter in the Fisher analysis.

		We parametrize the residues $R(\ell)$ of the observed and best-fit spectra as

		\begin{equation}
			\begin{aligned}
				R(\ell) & = D_{\ell}^{\rm TT, obs} - D_{\ell}^{\rm TT, best-fit} \\
						& = \sum_{i = 1}^{N_p} \delta A_i g_i^{\rm TT}(\ell) . 
			\end{aligned}
		\end{equation}

\begin{figure}[h]
			\includegraphics[width = 0.475\textwidth]{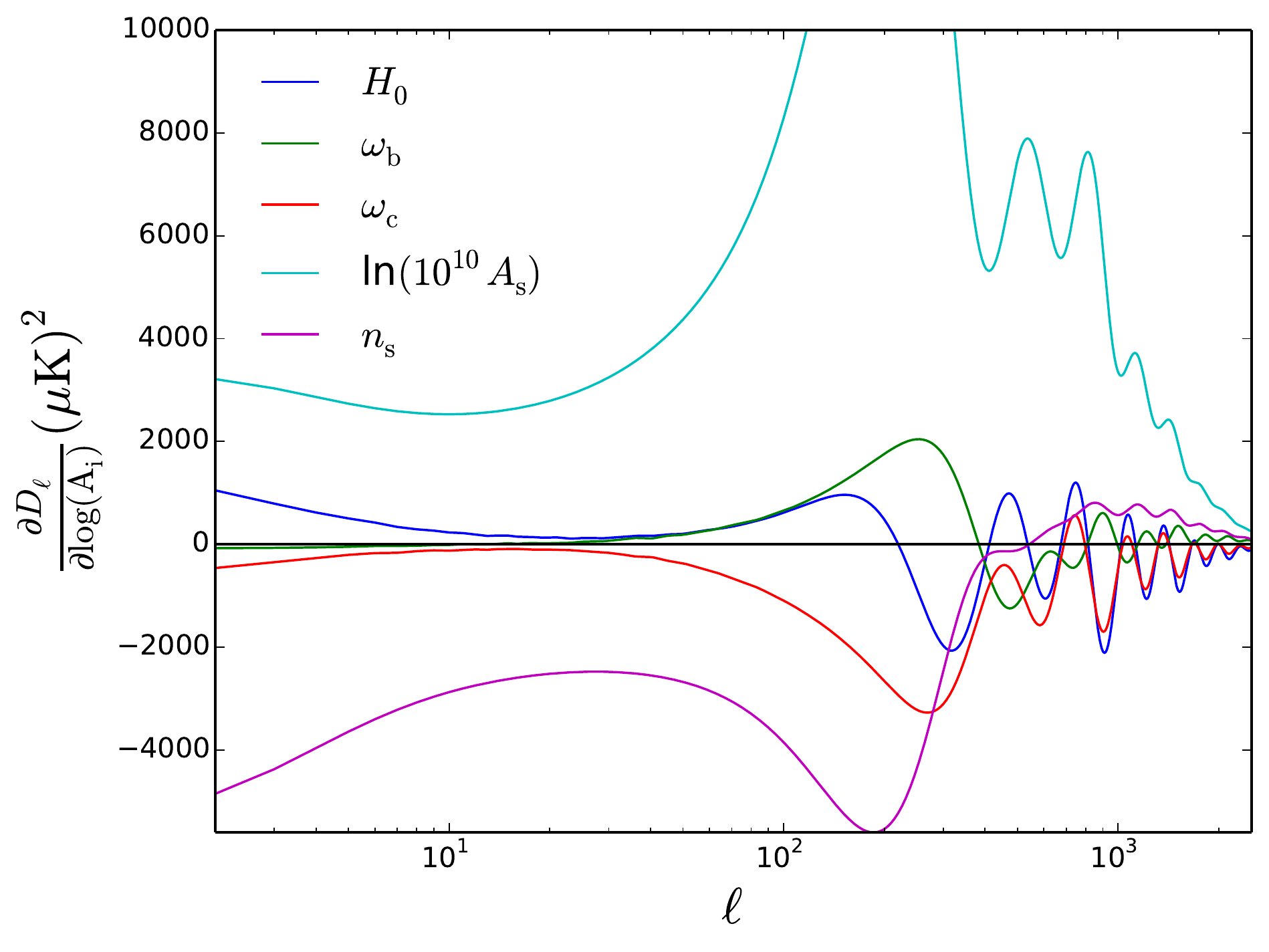}
			\caption{Shown here are the partial derivatives of the TT spectrum with respect to the cosmological parameters, $H_0$ (dark blue), $\omega_{\rm b}$ (green), $\omega_{\rm c}$ (red), ln($10^{10}A_{\rm s}$) (light blue) and $n_{\rm s}$ (pink). These were derived at the best-fit values obtained by setting $\tau = \tau_{\rm Pl}$, shown in Table \ref{best_fit_table}. }
			\label{CMB_partials}
		\end{figure}

		Here $N_p$ is the total number of parameters $A_i$, and 
		\begin{equation}
			g_i(\ell) = \frac{\partial D_{\ell}}{\partial A_i} . 
		\end{equation}
		where we have dropped the spectrum identifier TT. The partial derivatives $g_i(\ell)$ of the spectrum with respect to the cosmological parameters were determined by shifting the parameters by 1\% about their best-fit values and running the CLASS code to create the TT power spectrum for each shift. Therefore, $\Delta A_i = 0.01 A_i$ and the derivatives become:
		\begin{equation}
			g_i(\ell) = \frac{D_{\ell}(A_i + \Delta A_i) - D_{\ell}(A_i - \Delta A_i)}{2\Delta A_i} . 
		\end{equation}
		The choice of changing all parameters by 1\% is only somewhat arbitrary. We assume that this change is small enough that we are still in the linear regime, which validates the Fisher analysis and use of finite differences to numerically differentiate. Moreover, we assume a 1\% shift is large enough to ensure that the partial derivatives do not suffer significant numerical errors. These partial derivatives are shown in Fig.~\ref{CMB_partials}. 

\begin{figure}[h]
			\includegraphics[width = 0.475\textwidth]{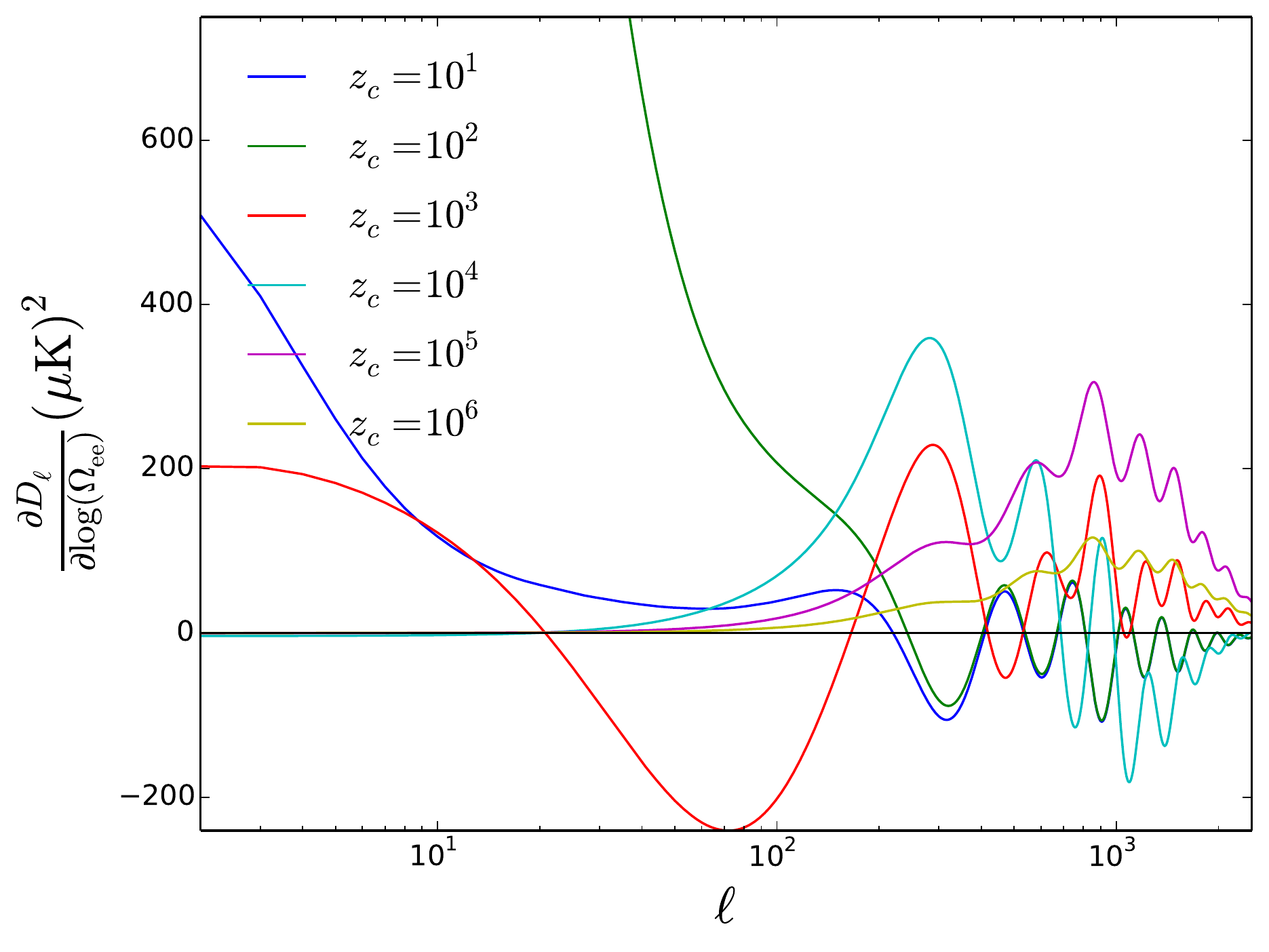}
			\caption{The partial derivatives of the TT spectrum with respect to \EE are shown here for various values of $z_c$. }
			\label{EE_partials}
		\end{figure}

		For the EE, the partials were determined as 
		\begin{equation}
			g_{\Omega_{\rm ee}}(\ell) = \frac{D_{\ell}(\Delta \Omega_{\rm ee}) - D_{\ell}(\Omega_{\rm ee} = 0)}{\Delta \Omega_{\rm ee}} , 
		\end{equation}
		where $\Delta$\EE is the value of \EE that moved the angular size $\theta_*$ of the sound horizon at the redshift of the CMB by 1\%. This value was found by recursively running CLASS for each $z_c$ until a $\Delta$\EE was found that moved $\theta_*$ by 1\% in either direction. The partial derivatives of $D_{\ell}^{\rm best-fit}$ with respect to \EE for various $z_c$'s are shown in Fig.~\ref{EE_partials}. 

		The Fisher matrix $F_{ij}$ is then given by
		\begin{equation}
			F_{ij} = \langle g_i , g_j \rangle , 
		\end{equation}
		where $\langle , \rangle$ denotes the inner product 
		\begin{equation}
			\langle g_i , g_j \rangle \equiv \sum_l \frac{g_i(\ell)  g_j(\ell)}{(\sigma_{D_{\ell}})^2} , 
		\end{equation}
		and $\sigma_{D_{\ell}}$ is the error on $D_{\ell}^{\rm obs}$. Hence, the analysis is limited by the error on the observed $D_{\ell}$'s. 

		The inverse Fisher matrix is then \cite{Coe:2009xf} 
		\begin{equation}
			(F^{-1})_{ij} = r_{ij} \sigma_i \sigma_j , 
		\end{equation}
		where $r_{ij}$ is the correlation coefficient between the parameters $A_i$ and $A_j$, and $\sigma_i$ and $\sigma_j$ are their respective errors.

	\subsection{Planck Data} \label{Sec_best_fit} \label{Sec_Pl_data}

		In their 2016 paper, Planck reports best-fit values for the TT + TE + EE + SIMLow (SimLow is based on low $\ell$ EE data) spectra combined \cite{Aghanim:2016yuo}. We begin by using these values for the cosmological parameters and for $\tau$; we label these as Planck-16. However, as we only use the Planck TT power spectrum for our analysis, the best-fit values for just the TT spectrum will be shifted from Planck-16 by some small amount. Therefore, we first do a Fisher analysis using just the TT spectrum and the cosmological parameters in order to find this new best fit. 

		The minimum-variance unbiased estimators are determined as 

		\begin{equation}
			\delta A_i = \sum_j (F^{-1})_{ij} \langle R(\ell) , g_j(\ell) \rangle , 
		\end{equation}
		where $\delta A_i$ quantifies the shift, relative to Planck-16, in the parameter $A_i$ that will fit just the TT data better.  We then check that the shifts in the parameters are all small compared with their $1\sigma$ errors and furthermore that the shift in 
		\begin{equation}
			\chi^2 = \sum_{\ell} \frac{R^2(\ell)}{(\sigma_{D_{\ell}})^2} = \langle R(\ell) , R(\ell) \rangle 
		\end{equation}
		is insignificant.  We thus check that
		\begin{equation} \label{small_partial_chi}
			\sigma_i \frac{\partial \chi^2}{\partial A_i} = -2 \sigma_i \langle R(\ell) , g_i(\ell) \rangle \ll 1 . 
		\end{equation} 
		Doing so, we begin our investigation of the effects of exotic energy with a baseline $\Lambda$CDM model that provides the best fit to the TT data that we use and that is consistent, within errors, with the best-fit CMB values obtained from the full Planck-16 analysis. The values adopted for the cosmological parameters + $\tau$ for our subsequent analysis are shown in Table \ref{best_fit_table}. 

		As errors on higher $D_{\ell}$'s are correlated \cite{Ade:2013kta}, we use binned data for $\ell \geq 30$. The bin size is 30 for all but the last bin which spans $2490 \leq \ell \leq 2508$. The correlation between errors on $D_{\ell}$'s from different bins is then diminished. In all, we use $2 \leq \ell \leq 2508$ for the analysis.

		\begin{table*}[t] 
			\centering
			\small 
			\setlength{\tabcolsep}{12pt} 
			\begin{tabular}{c | *{5}{c}} 
					& Planck-16					& $\tau = \tau_{\rm Pl}$	& $\tau = \tau_{\rm Pl} + 2\sigma_{\tau,\rm Pl}$	& $\tau = \tau_{\rm Pl} + 5\sigma_{\tau,\rm Pl}$ \\
				\hline \hline
$100h$				& 66.93	$\pm$ 0.62			& 67.99749				& 68.28782									& 68.77709 \\
$\omega_{\rm b}$	& 0.02218 $\pm$ 0.00015		& 0.02240				& 0.02244									& 0.02251 \\
$\omega_{\rm c}$	& 0.1205 $\pm$ 0.0014		& 0.11970				& 0.11906									& 0.11799 \\
$\tau$				& 0.0596 $\pm$ 0.0089		& 0.0596				& 0.0774									& 0.1041 \\
$ln10^{10}A_{\rm s}$& 3.056 $\pm$ 0.018			& 3.05576				& 3.08972 									& 3.14024 \\
$n_{\rm s}$			& 0.9619 $\pm$ 0.0045		& 0.96453 				& 0.96599 									& 0.96862 \\
$\chi^2_{\rm red}$ 	& 0.9271					& 0.7652 				& 0.7471 									& 0.7322 \\
			\end{tabular}
			\caption{The values of the cosmological parameters and reionization optical depth $\tau$ used as the best-fit values with no exotic energy (EE) are shown alongside the Planck values. We also show the reduced $\chi^2$ for the TT power spectrum for these values. }
			\label{best_fit_table}
		\end{table*}

\section{Constraints on the EE} \label{sec_constraints}
	
	Adding the EE will shift all parameters by some amount, which can be expressed in terms of $\chi^2$ and the errors on the parameters as 
	\begin{equation}
		\delta A_i = - \frac{1}{2} \sum_j r_{ij} \sigma_i \sigma_j \frac{\partial \chi^2}{\partial A_j} . 
	\end{equation}

	The quantity $\sigma_j (\partial \chi^2/\partial A_j)$ is small at the best-fit value for the cosmological parameters and the correlation coefficients are such that $|r_{ij}| \leq 1$. This makes the shift in any parameter $\delta A_i$ due to any of the cosmological parameters much smaller than the error $\sigma_i$ on $A_i$. Therefore, all significant shifts are due to the EE,  
	\begin{equation}
		\delta A_i \simeq - \frac{1}{2} (F^{-1})_{i,\Omega_{\rm ee}} \frac{\partial \chi^2}{\partial \Omega_{\rm ee}} . 
	\end{equation}
	For the EE, this shift looks like 
	\begin{equation}
		\delta \Omega_{\rm ee} \simeq - \frac{1}{2} (F^{-1})_{\Omega_{\rm ee},\Omega_{\rm ee}} \frac{\partial \chi^2}{\partial \Omega_{\rm ee}} . 
	\end{equation} 
	Therefore, the shift in parameter $A_i$ induced by a change $\Delta $\EE from its baseline value \EE $=0$ is
	\begin{equation}
		\delta A_i (z_c) \simeq \frac{(F^{-1})_{i,\Omega_{\rm ee}}}{(F^{-1})_{\Omega_{\rm ee},\Omega_{\rm ee}}} \delta \Omega_{\rm ee} , 
	\end{equation} 
	where $\delta A_i$'s are now a function of the critical redshift $z_c$. 

	Below we do the following for several values of $\tau$:  (1) We first determine the values of the five other cosmological parameters that provide the best fit to the TT data we use; (2) we then add \EE as a sixth parameter to the Fisher analysis and determine (a) the best-fit value of \EE; (b) the $1\sigma$ error to \EE; and (c) the shifts induced by \EE to the cosmological parameters and record specifically the shift in $H_0$. We provide results as a function of $10 \lesssim z_c \lesssim 10^6$. (3) We look in each case to see whether the introduction of \EE improves the fit to the TT data by a statistically significant amount.  In no case do we find evidence that the TT data prefers a nonzero value of \EE and thus derive in each case only upper limits to $\Omega_{\rm ee}$.  

	\subsection{Fixing $\tau = \tau_{\rm Pl}$} \label{sec_Pl_tau}

\begin{figure}
				\includegraphics[width = 0.475\textwidth]{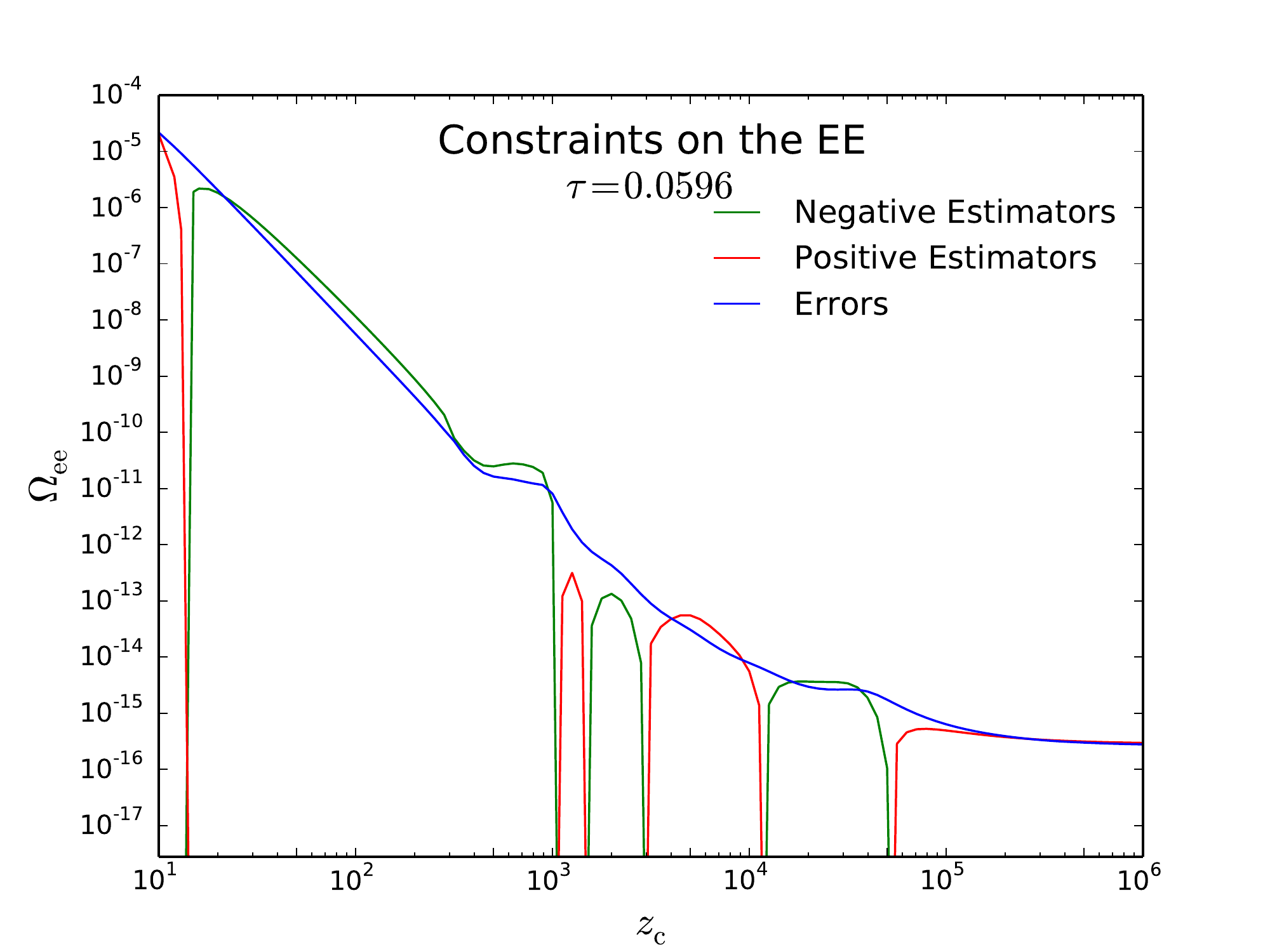}
				\caption{The best-fit values and errors on \EE are shown here. The optical depth $\tau$ was fixed at the best-fit Planck-16 value to obtain these constrains. }
				\label{Pl_tau_EE_const}
			\end{figure}

			We begin by considering the current Planck central value $\tau=0.0596$.  The constraints to $\Omega_{\rm ee}$ are then shown in Fig.~\ref{Pl_tau_EE_const} as a function of the critical redshift $z_c$.  Also shown there is the $1\sigma$ error to $\Omega_{\rm ee}$.  The best-fit value of $\Omega_{\rm ee}$ is (unphysically) negative for some $z_c$, but for no value of $z_c$ does the preferred value depart from the null result by a statistically-significant amount. This remains true for all our constraints on \EE for various values of $\tau$. 

			For $\tau=0.0596$, the largest allowable EE-induced increase in the best-fit value of the Hubble parameter is 0.22 km~s$^{-1}$~Mpc$^{-1}$, at a critical redshift $z_c \simeq 10000$.  This is a small fraction of the Planck $1\sigma$ error (roughly 0.6 km~s$^{-1}$~Mpc$^{-1}$) to $H_0$, so does not do much in the way of relieving the CMB/local-measurement tension.  The introduction of $\Omega_{\rm ee}$ to the Fisher analysis increases the error to $H_0$, to roughly 1.2 km~s$^{-1}$~Mpc$^{-1}$, and so may go some way toward alleviating the tension.

			\begin{figure}
				\includegraphics[width = 0.475\textwidth]{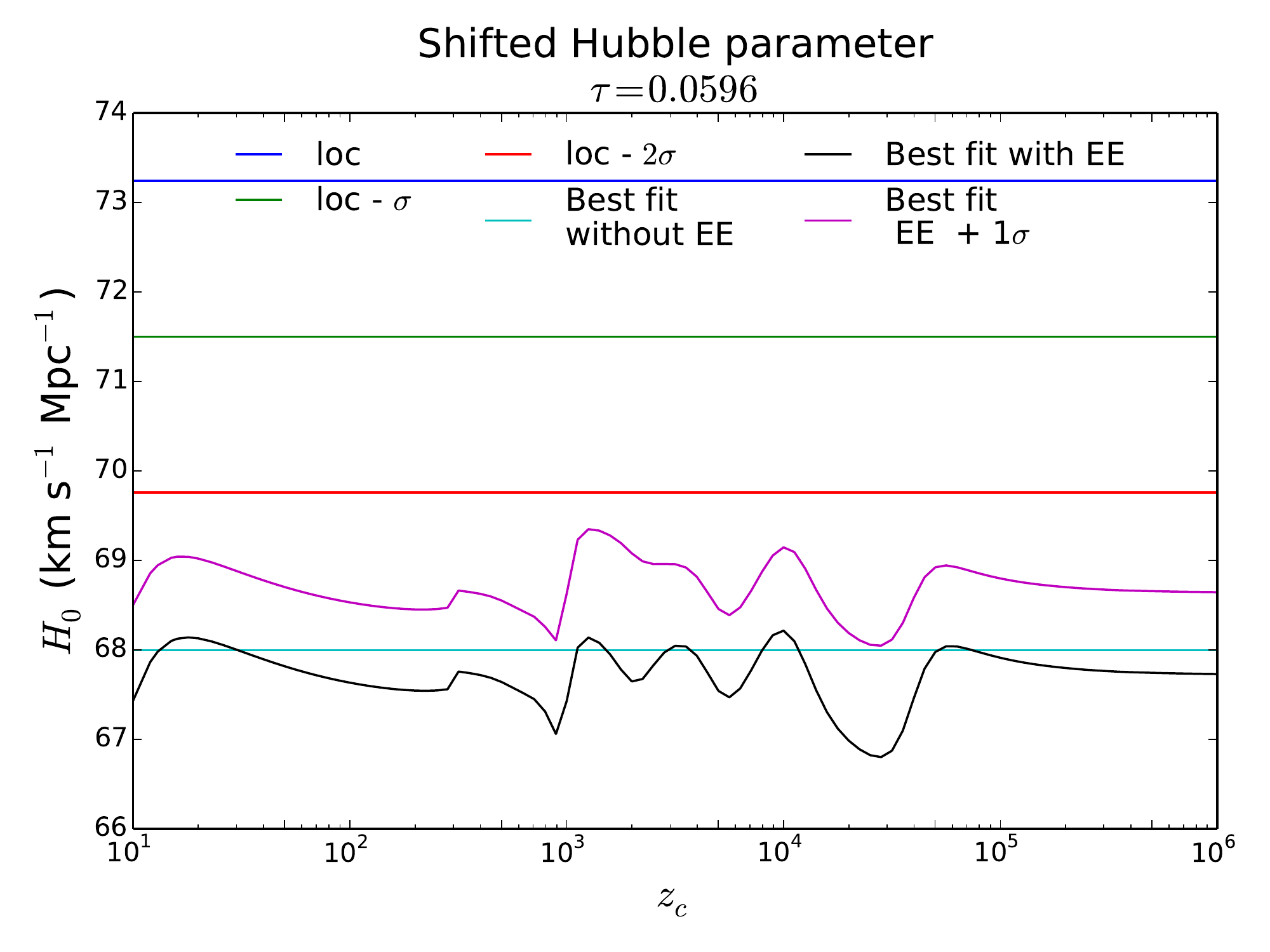}
				\caption{Shown here are the best-fit values of the Hubble parameter $H_0$ and its $1\sigma$ upper limit obtained by including EEs in the fit to the Planck temperature power spectrum.  We also show the central value obtained from local measurements in \cite{Riess:2016jrr} as well as the values that are $1\sigma$ and $2\sigma$ lower than the best fit. }
				\label{Pl_tau_H0}
			\end{figure}

	\subsection{Fixing $\tau = \tau_{\rm Pl} + 2\sigma_{\tau,\rm Pl}$} \label{sec_2sig}

			Next we fix $\tau$ at its Planck-16 $2\sigma$ upper limit. The TT spectrum prefers a larger value of $\tau$ \cite{Ade:2013zuv}. Therefore, the reduced $\chi^2$ is slightly smaller in this case, and smaller still when we fix $\tau$ at its $5\sigma$ Planck-16 value, as seen from Table~\ref{best_fit_table}. 

			\begin{figure}
				\includegraphics[width = 0.475\textwidth]{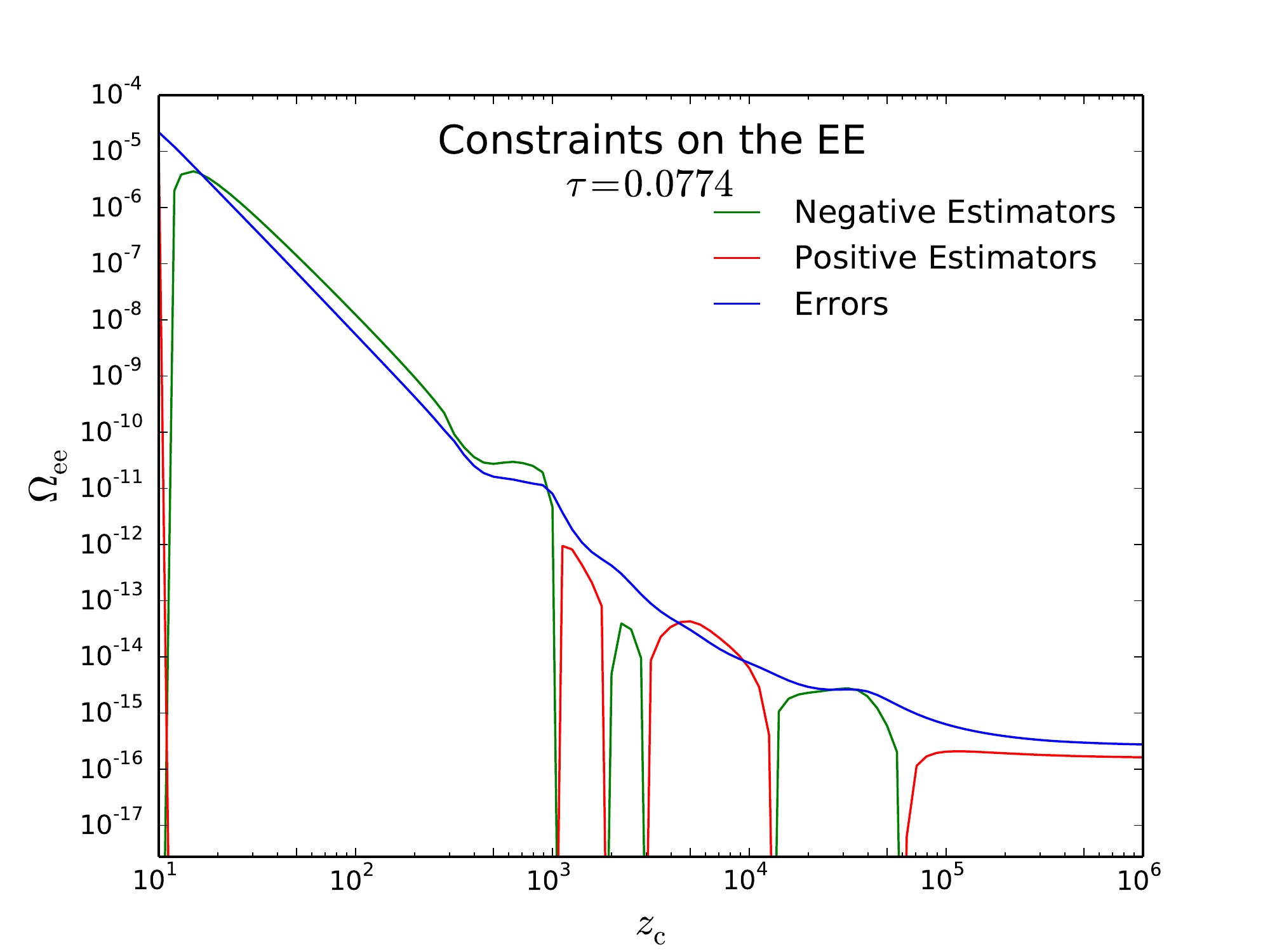}
				\caption{The best-fit values and errors on \EE are shown for various critical redshifts of the EE. We fix $\tau$ at $\tau_{\rm Pl} + 2\sigma_{\tau,\rm Pl}$.  }
				\label{Pl_tau_2sig_EE_const}
			\end{figure}

			\begin{figure}
				\includegraphics[width = 0.475\textwidth]{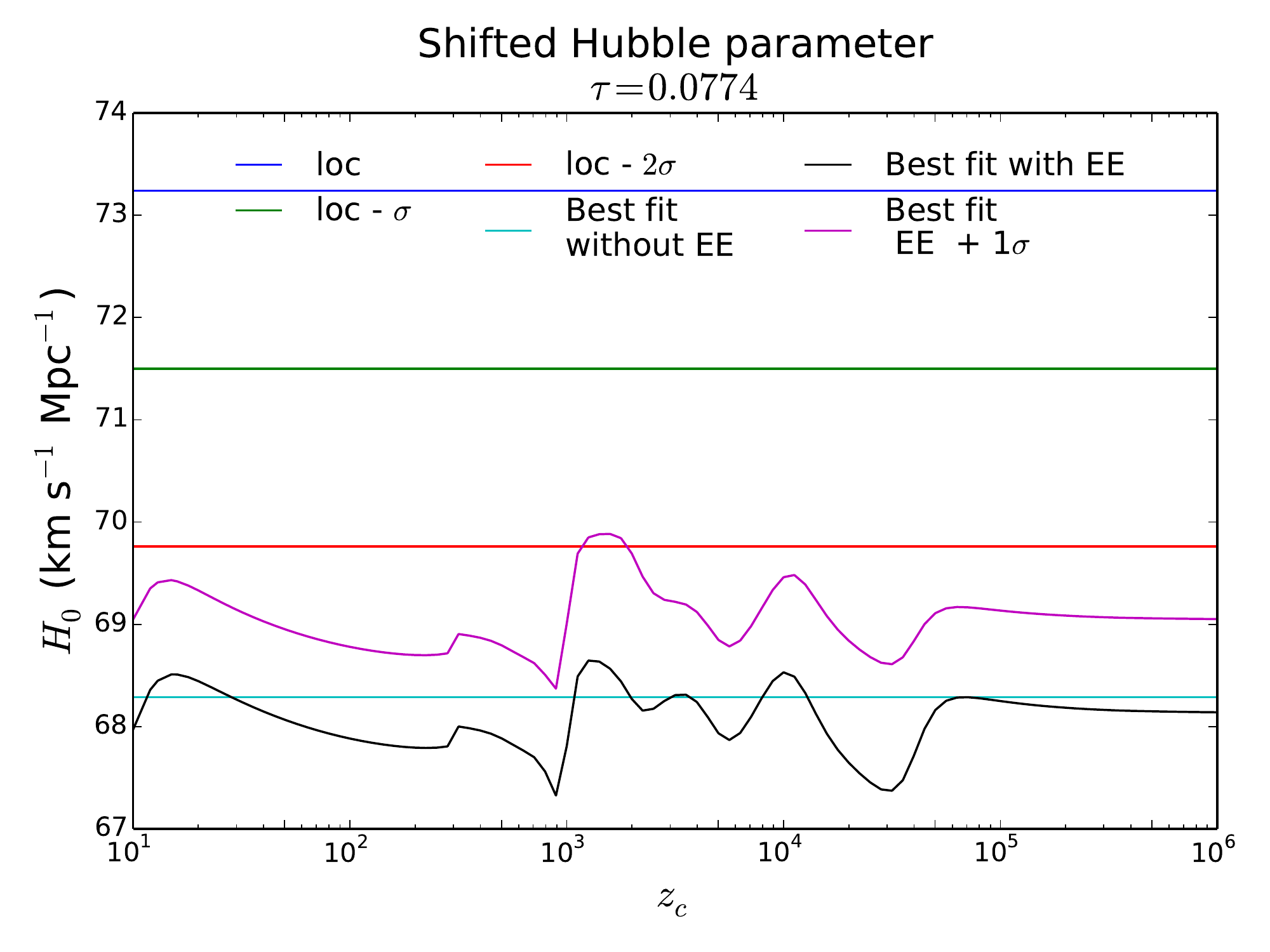}
				\caption{The best-fit and best-fit + $1\sigma$ values for $H_0$ (in km s$^{-1}$ Mpc$^{-1}$) are shown along with its local measurement at various $\sigma$.}
				\label{Pl_tau_2sig_H0}
			\end{figure}

			The constraints on \EE are shown in Fig.~\ref{Pl_tau_2sig_EE_const}. We find that the errors on \EE are essentially the same between our analyses at various values of $\tau$. The blue line in Fig.~\ref{Pl_tau_2sig_EE_const} hence offers a visual reference to comparing constrains on \EE for various $\tau$. 

			The change brought about in the Hubble parameter for $\tau = 0.0774$ is shown in Fig.~\ref{Pl_tau_2sig_H0}. The best-fit value of $H_0$ increases at most by 0.36 km s$^{-1}$ Mpc$^{-1}$ ($z_c = 1259$), its 1$\sigma$ value increasing at most by 1.6 km s$^{-1}$ Mpc$^{-1}$ ($z_c = 1585$). The total increase in the Hubble parameter for $\tau = 0.0774$ is twofold. Firstly, the EE is capable of inducing a greater positive shift in $H_0$ as compared to $\tau = \tau_{\rm Pl}$. Secondly, for higher $\tau$, a larger best-fit value of $H_0$ without any EE is preferred, as can be seen from Table 1. Consequently, although the Hubble tension is not resolved, $H_0$ is pushed closer to its local measurement.

		\subsection{Fixing $\tau = \tau_{\rm Pl} + 5\sigma_{\tau,\rm Pl}$} \label{sec_5sig}

			The results from fixing $\tau$ at its Planck-16 and $2\sigma$ values hint that perhaps a higher value of $\tau$ will allow the EE to fully resolve the Hubble tension. Therefore, in this Section we explore what happens if $\tau$ for some reason departs by $5\sigma$ from its best-fit value. The best-fit values adopted in this section are shown in Table \ref{best_fit_table}. 

\begin{figure}
				\includegraphics[width = 0.475\textwidth]{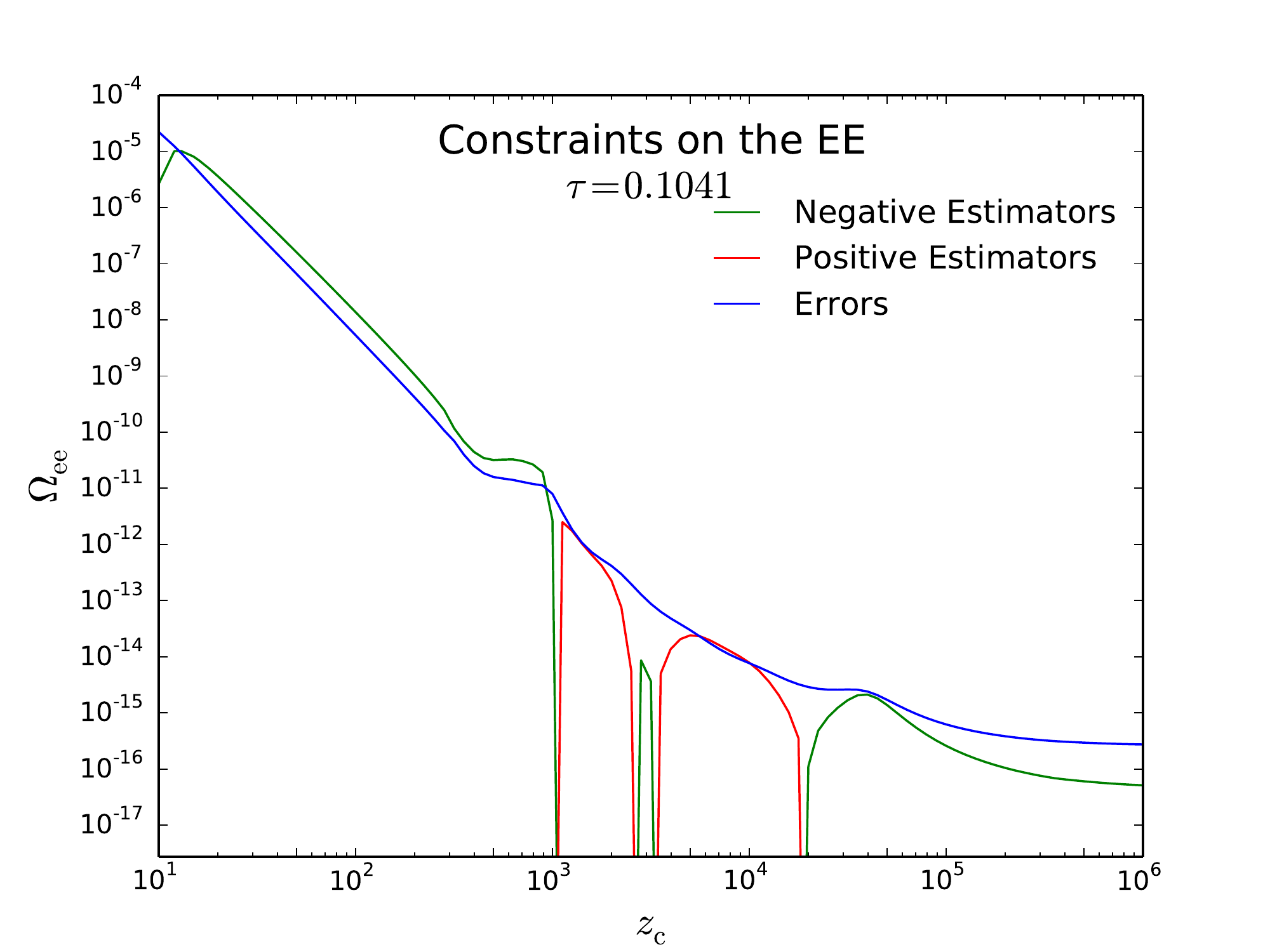}
				\caption{The best-fit values and errors on \EE for various $z_c$ are shown for $\tau$ fixed at $\tau_{\rm Pl} + 5\sigma_{\tau,\rm Pl}$. }
				\label{Pl_tau_5sig_EE_const}
			\end{figure}

\begin{figure}
				\includegraphics[width = 0.475\textwidth]{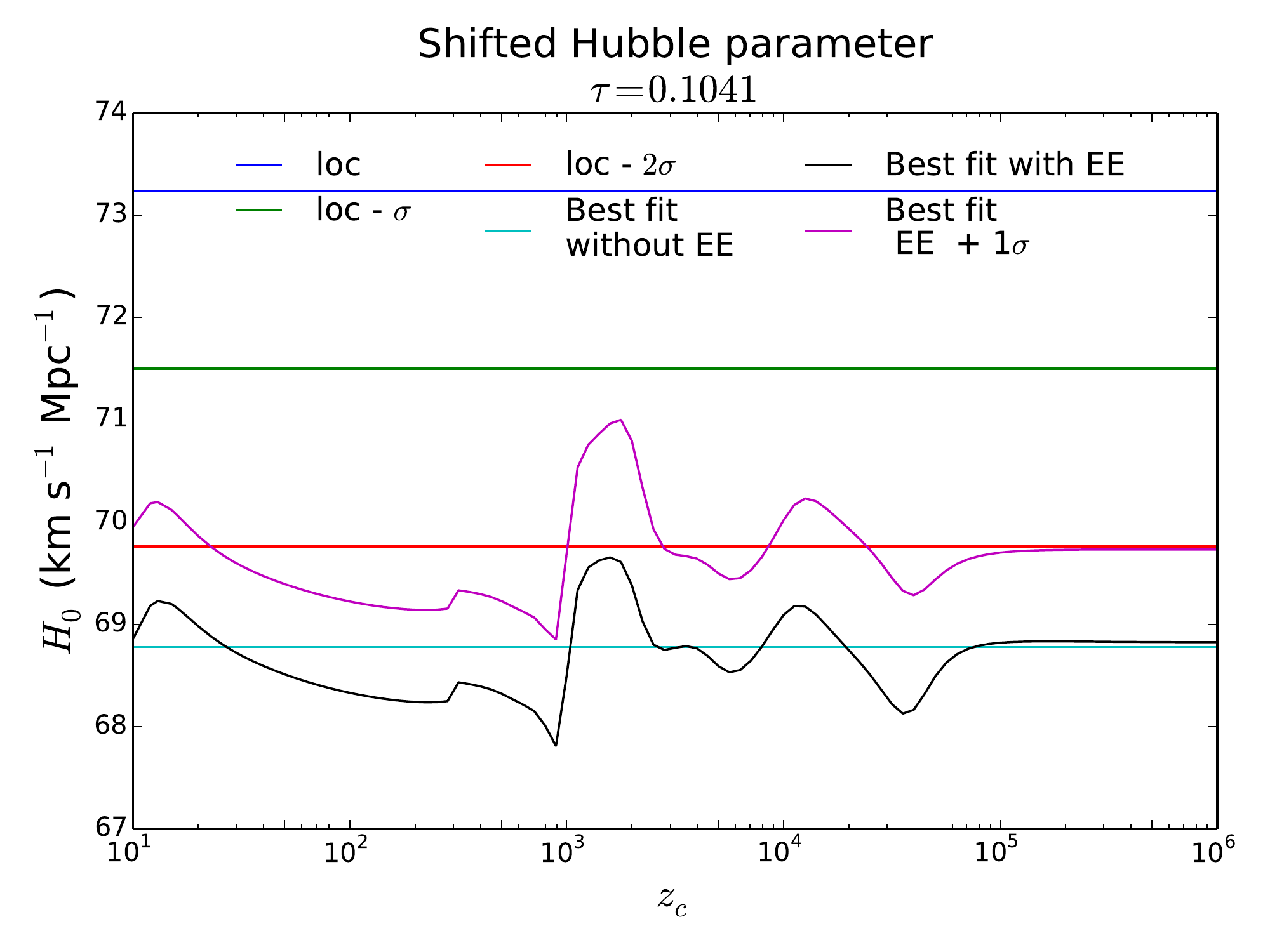}
				\caption{The best-fit vales of $H_0$ (in km s$^{-1}$ Mpc$^{-1}$) are shown with their $1\sigma$ errors. The local measurement is also shown at various $\sigma$.}
				\label{Pl_tau_5sig_H0}
			\end{figure}

			The constraints we obtain on \EE are shown in Fig.~\ref{Pl_tau_5sig_EE_const}. The change in the Hubble parameter is shown in Fig.~\ref{Pl_tau_5sig_H0}. 

			Although fixing $\tau$ at $5\sigma$ does not entirely eliminate the discrepancy, $H_0$ is increased by a greater amount as compared to Fig.~\ref{Pl_tau_2sig_H0}. For some $z_c$, it is increased to within the $2\sigma_{\rm loc}$ range of the locally measured Hubble parameter $H_{0, \rm loc}$. The greatest increase in the best-fit value of $H_0$ is 0.88 km s$^{-1}$ Mpc$^{-1}$ ($z_c = 1585$), in its $1\sigma$ value is 2.22 km s$^{-1}$ Mpc$^{-1}$ ($z_c = 1779$). 

			\begin{figure*}
				\includegraphics[width = 1\textwidth]{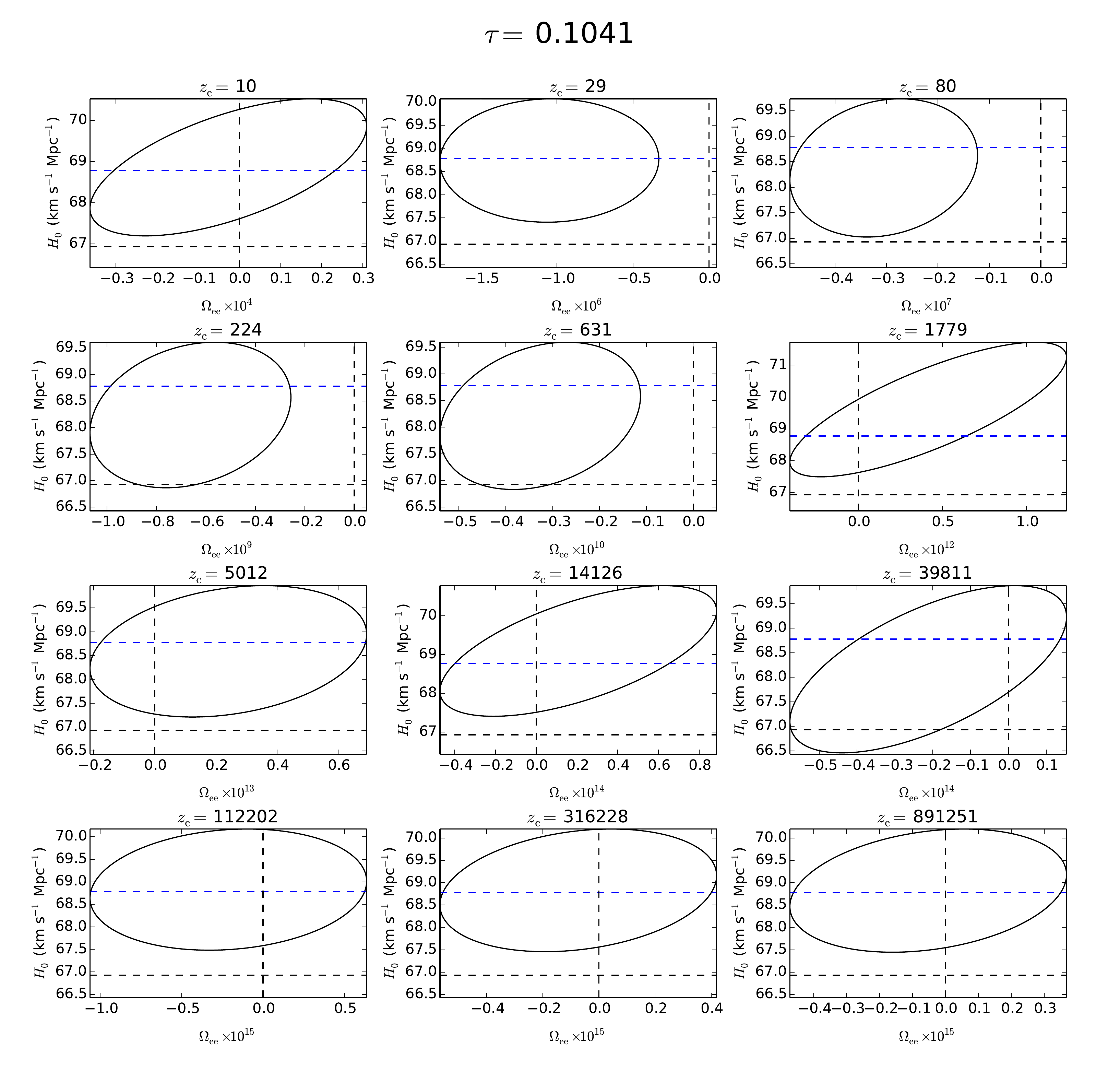}
				\caption{We plot the $1\sigma$ likelihood contours for the Hubble parameter against \EE for various critical redshifts, covering the range of critical redshifts that we probe. We fix $\tau = \tau_{\rm Pl} + 5\sigma_{\tau,\rm Pl}$ for these. In each plot, the Planck-16 values for both parameters are marked by the horizontal and vertical dashed black lines. The dashed blue lines mark the value for the best-fit Hubble parameter for just the TT spectrum without EEs. Negative values of \EE are unphysical but allowed in our analysis. Estimators of \EE are consistent with zero within $\sim 2\sigma$. }
				\label{EE_H0_5sig_contours}
			\end{figure*}

			We plot $1\sigma$ likelihood ellipses for $H_0$ and \EE in Fig.~\ref{EE_H0_5sig_contours}. For local extrema in the shifts in $H_0$, a higher correlation between $H_0$ and \EE can be seen in the ellipses. While for critical redshifts that leave $H_0$ unchanged, there is little correlation between $H_0$ and $\Omega_{\rm ee}$. The Planck-16 values are always within $\sim 2\sigma$ ellipses and all the \EE estimators are consistent with the null result.

\section{Conclusions} \label{sec_conclusion}
	
	We consider a simple exotic energy density that provides a small perturbation to standard $\Lambda$CDM. The EE behaves like a cosmological constant until some critical redshift $z_c$, then decays away as $a^{-6}$. We investigate whether such an EE can alleviate the Hubble tension and find constraints on the maximum fractional energy density \EE today, that this field can have by doing a Fisher analysis on the Planck TT power spectrum. 

	In our analysis, we find that the value of $\tau$ places a strong constraint on the preferred value of \EE as well as the extent to which it can mitigate the Hubble tension. A larger value of $\tau$ leads, with EE, to a larger best-fit value of $H_0$.

	In order for the best-fit value of $H_0$ for a $\Lambda$CMD + EE universe to coincide with the local measurement, a value of $\tau$ greater than its 5$\sigma$ Planck-16 value is required. (Such a large value of $\tau$ is consistent with that obtained by the WMAP 9-year results, $\tau_{\rm WMAP} = 0.088 \pm 0.014$ \cite{Hinshaw:2012aka}.) If we fix $\tau$ at its Planck-16 best-fit and $2\sigma$ values, the tension is not altogether resolved, however, $H_0$ is shifted up closer to its local value. This is largely due to the error on $H_0$ increasing on the addition of the EE. Increasing $\tau$ and allowing for such an EE is indeed capable of alleviating the Hubble tension. 

	The Hubble tension between local measurements and the Planck data has been studied before by Ref.~\cite{Riess:2016jrr, Bernal:2016gxb, DiValentino:2016hlg, Grandis:2016fwl, Archidiacono:2016kkh, Umilta:2015cta, Huang:2015srv, Ade:2015rim, Joudaki:2016kym}. Altering the effective number of neutrino species $N_{\rm eff}$ \cite{Riess:2016jrr} and allowing the equation of state parameter of dark energy $w$ to vary with time \cite{Joudaki:2016kym} have been investigated as solutions to the Hubble tension (although variable $w$ may introduce more tensions, eg. with BAO \cite{Joudaki:2016kym}). The correlation between $H_0$ and $N_{\rm eff}$ as well as that between $H_0$ and variable $w$ is stronger than that between $H_0$ and the EE and they may be better candidates for diminishing the Hubble tension. 

	Furthermore, Ref.~\cite{Riess:2016jrr,Grandis:2016fwl,Addison:2015wyg} suggest unresolved systematics in Planck data may be the cause of the tension. In particular, Ref.~\cite{Addison:2015wyg} suggests that Planck multipoles $\ell \geq 1000$ may suffer systematic errors. Excluding $\ell \geq 1000$ data not only significantly reduces the Hubble tension, but would also allow more room for early dark energy. However, Ref.~\cite{Aghanim:2016sns} finds inconsistencies between high and low multipoles in Planck data statistically insignificant. 

	\begin{figure*}
		\includegraphics[width = 1\textwidth]{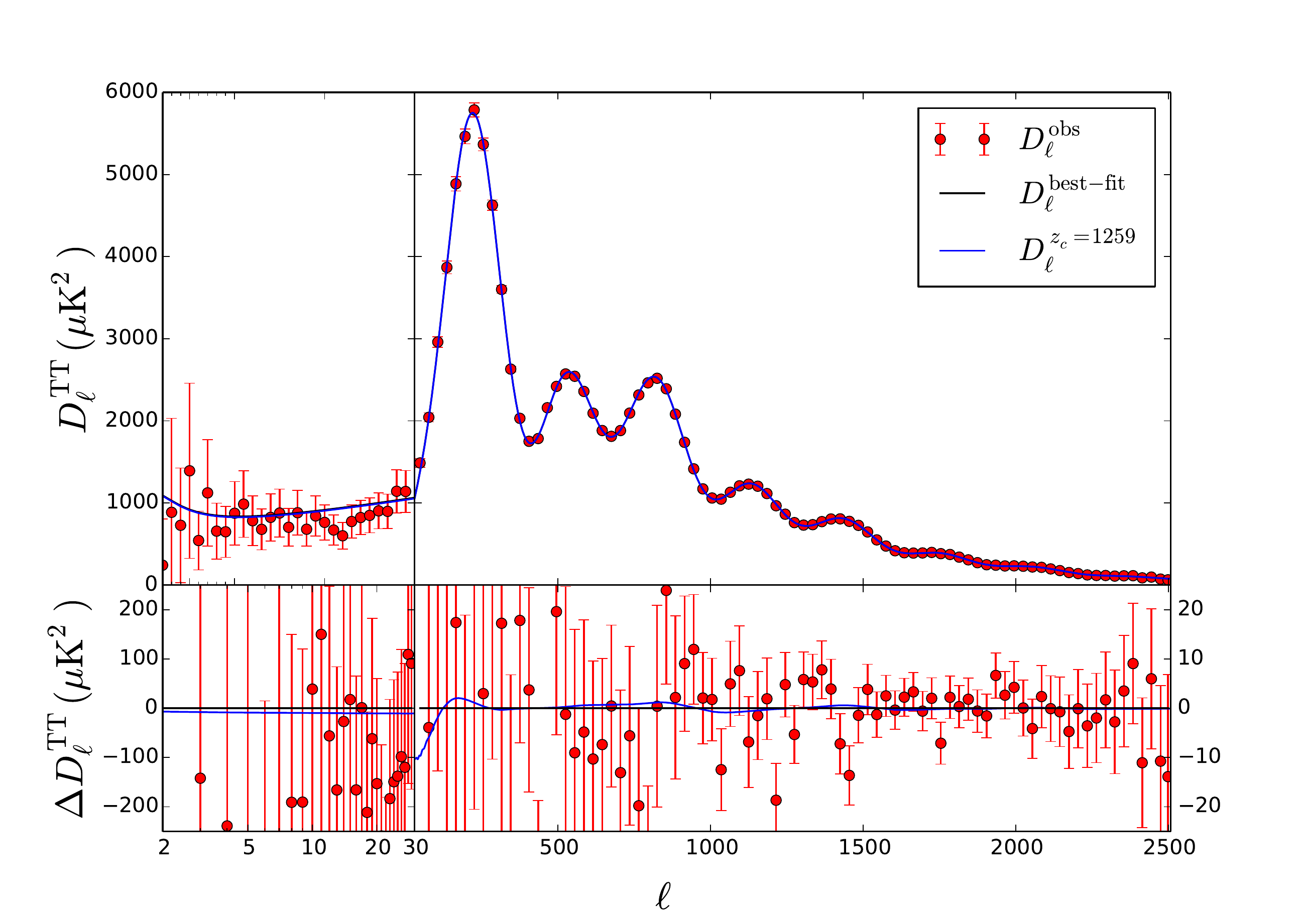}
		\caption{We plot two best fits (blue and black) for the Planck temperature angular power spectrum for $\tau = \tau_{\rm Pl} + 2\sigma_{\tau,\rm Pl}$, and the Planck data (red). In black is the best fit without any EE. In blue we plot the best fit including an EE with $z_c = 1259$. This is the EE that increases the best-fit Hubble parameter the most. In the lower panel, we subtract $D_{\ell}^{\rm best-fit}$ from all three spectra and plot the residues. The bottom left and bottom right panels are scaled differently such that the residues may be more easily distinguishable. }
		\label{Residues_2sig}
	\end{figure*}

	Adding the EE to $\Lambda$CDM, the cosmological parameters shift to accommodate the EE. The reduced $\chi^2$ for the TT spectrum at their new best fit is not significantly changed. All changes in $\chi^2_{\rm red}$ are approximately an order smaller than the error on it. In Fig.~\ref{Residues_2sig}, we plot the best-fit spectra without any EE and that with the EE which increases the best-fit vale of $H_0$ the most, for $\tau = 0.0774$. From the residues in the lower panel shown therein, it can be seen that the addition of the EE leaves the TT spectrum, and hence the reduced $\chi^2$s, largely unaltered. Therefore, current data does not favor with statistical significance the addition of the EE to $\Lambda$CDM.

\begin{figure}
		\includegraphics[width = 0.475\textwidth]{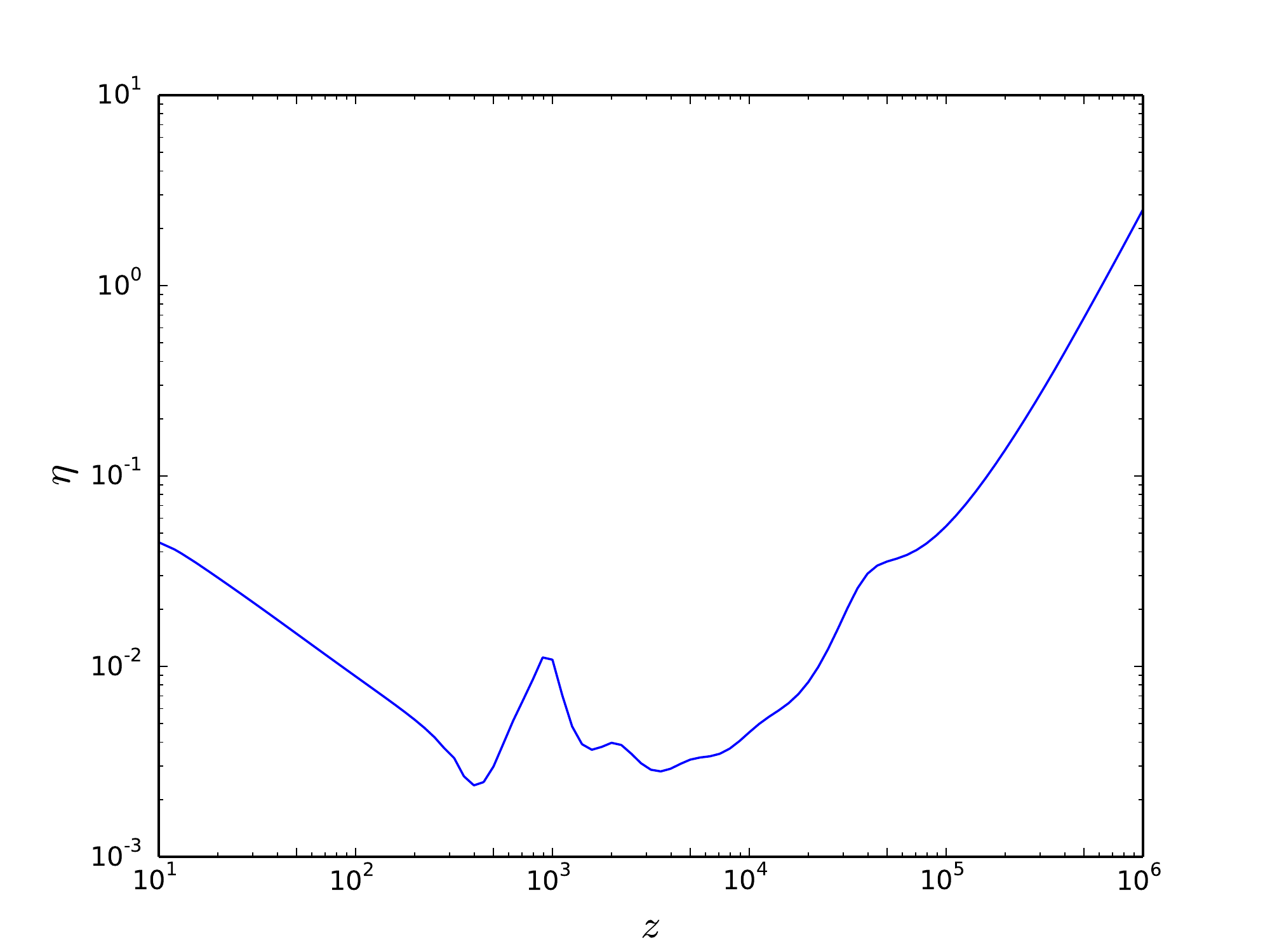}
		\caption{The exotic energy density at its critical redshift $\rho_{\rm ee}(z_c)$ is plot as a fraction of the total energy density $\rho_{\Lambda\rm CMD}(z_c)$ of a pure $\Lambda$CDM universe for a range of redshifts. This fraction is within an order-unity factor of the greatest contribution of the EE to the energy density of the Universe. This plot was made for $\tau = \tau_{\rm Pl}$ and allowing \EE $= \sigma_{\Omega_{\rm ee}}$. }
		\label{fig_EE_max_contr}
	\end{figure}

	This EE was motivated from axion-like fields that may explain dark energy \cite{Kamionkowski:2014zda}. The exotic energy considered here contributes its most to the total energy density of the Universe close to its critical redshift, forming its greatest fraction of the total energy density of the Universe. In Fig.~\ref{fig_EE_max_contr} we plot this fraction $\eta = \rho_{\rm ee}(z_c)/\rho_{\Lambda \rm CDM}(z_c)$ of the total energy density of a pure $\Lambda$CDM universe that early exotic dark energy can from, as a function of redshift, according to our constraints on $\Omega_{\rm ee}$. For extremely high redshifts, the TT spectrum allows dark energy to have a larger energy density than that in a $\Lambda$CDM universe as long as it quickly redshifts away. This can also be seen from Fig.~\ref{EE_behaviour}, where the EE with the greatest critical redshift has a higher energy density than radiation just before it decays. Closer to recombination, the greatest contribution of early dark energy is constraint to be $\lesssim 2\%$ of the total energy density in a $\Lambda$CDM universe. This result is consistent with constraints on other early dark energy models obtained through Monte Carlo analyses \cite{Doran:2006kp,Pettorino:2013ia,Calabrese:2011hg} that found upper limits of 4-5\%. 

	The constraints presented here on \EE can be improved by more computationally heavy approaches such as including polarization data in the analysis or by doing a full MCMC on the 6 dimensional parameter space for each $z_c$ considered. However, our simpler approach allows us to constrain an early dark energy model on a level consistent with a full MCMC analysis, and show that it is capable of increasing the value of the Hubble parameter. We conclude that adding an exotic energy, such as the one considered here, to $\Lambda$CDM may form a part of the solution to the Hubble tension if a higher optical depth to reionization is allowed. If the Hubble tension persists with a 1\% measurement of the local value of $H_0$, then it may be useful to revisit the exotic-energy model considered here.

\section{Acknowledgments}

	We thank Julian Mu\~noz and Daniel Pfeffer for helpful discussions, and Adam Riess for useful comments on an earlier draft. This work was supported by NSF Grant No. 0244990, NASA NNX15AB18G, the John Templeton Foundation, and the Simons Foundation. 

\bibliography{bib}

\end{document}